\documentstyle[sprocl]{article}

\newcommand{\nwc}{\newcommand}
\def\gtrsim{\mathrel{\relax{\raisebox{3pt}{$\mathord{>}$} \kern-0.75em
\raisebox{-2pt}{$\sim$}}}}
\def\lesssim{\mathrel{\relax{\raisebox{3pt}{$\mathord{<}$} \kern-0.75em
\raisebox{-2pt}{$\sim$}}}}
\newcommand{\be}{\begin{equation}}
\newcommand{\ee}{\end{equation}}

\newcommand{\text}{\textstyle}
\newcommand{\scr}{\scriptstyle}
\newcommand{\sscr}{\scriptscriptstyle}
\newcommand{\refs}[1]{(\ref{#1})}
\nwc{\ra}{\rightarrow}
\nwc{\pa}{\partial}
\def\a{\alpha}
\def\b{\beta}
\def\g{\gamma}
\def\vf{\varphi}
\def\l{\lambda}
\def\m{\mu}
\def\n{\nu}
\def\o{\omega}
\def\D{\Delta}
\def\F{\Phi}
\def\Ricci{{\cal R}}
\def\cl{{\cal L}}
\def\co{{\cal O}}
\def\bk{{\bf k}}

\begin{document}

\title{BACKREACTIONS IN SUPERINFLATIONARY COSMOLOGIES}
\author{ R. POPPE }
\address{Physik Department, Technische Universit\"at M\"unchen, \\ 85747
  Garching, Germany}
\author{ Z. LALAK }
\address{Institute of Theoretical Physics, University of Warsaw, \\ Ho\.za 69,
  00--681 Warsaw, Poland}

\maketitle\abstracts{ We examine Jordan--Brans--Dicke theories with a linear
  potential for the scalar field by means of a stochastic approach. The
  backreaction of the scalar field fluctuations on the classical background is
  described. The analysis is concentrated on those values of the
  Jordan--Brans--Dicke parameter $\o$ which predict a superinflationary
  pre--big--bang branch. The graceful exit problem present in these theories
  is discussed. }

In scalar tensor theories of gravity, such as Jordan--Brans--Dicke (JBD)
theories~\cite{brans}, the gravitational coupling constant is replaced by a
dynamical scalar field $\F = m_{Pl}^{2} = G^{-1}$. They are parameterised by a
dimensionless kinetic coupling parameter $\o$ and run into Einstein gravity in
the limit $\o\ra\infty$.

Theories of gravity arising from Kaluza--Klein theories or supergravity may
appear after compactification to four dimensions as a JBD theory with
$\o\simeq\co (1)$~\cite{freund}. The universal part of the low energy
effective string action in four dimensions can be viewed as a JBD theory as
well, with the specific choice $\o =-1$. For negative $\o $, the theories
predict a {\em superinflationary} phase defined for all times $t<0$ (the
curvature singularity at $t=0$ is identified with the big bang), hence the
name pre--big--bang branch. It is driven entirely by the non standard kinetic
term of the scalar field in the JBD frame~\cite{levin}.

Despite some attractive features of the superinflationary cosmology a
natu\-ral exit out of this phase into an ordinary radiation or matter
dominated universe is not easily accomplished~\cite{bru_ven}. Proofs were
given that on the classical level such a transition cannot
occur~\cite{kal_mad}, even in the case, when a potential for the field is
present. On the other hand, there are indications that a graceful exit can be
achieved by higher curvature terms and string loop
corrections~\cite{brustein-madden}. These terms get important in the strongly
coupled regime close to the singularity and quantum effects are no longer
negligible there. In the spirit of a QFT on a curved semiclassical background,
we shall examine the stochastic properties of the fluctuating degrees of
freedom and their backreaction on the homogeneous background. We can merely
report the results here and present the complete analysis
elsewhere~\cite{lalak-poppe}.

\noindent
The JBD action including a scalar potential and matter contributions reads
\be
S=\frac{1}{16 \pi }\int d^{4}x\;\sqrt{-g}\,\left[\;\F\,\Ricci - 
\o\,\frac{\pa_{\m}\F\,\pa^{\m}\F}{\F} - V(\F) + 16\pi\cl_{m}
\;\right] \ ,
\label{action} 
\ee
where $\Ricci$ is the Ricci scalar. From now on we ignore all contributions
from ordinary matter and assume a flat FRW background metric characterised by
a single scale factor $a(t)$. The equations of motion admit the following
solutions for the homogeneous background quantities if no potential is present,
\be
\F_{0}\, (t) \sim |\, t\, |^{\text{r_{\sscr{\pm}}(\o)}}
\qquad\quad\mbox{and}\qquad\quad 
a (t) \sim |\, t\, |^{\text{q_{\sscr{\pm}}(\o)}} \ .
\label{V0solutions}
\ee
The subscripts in the exponents $r_{\sscr{\pm}}$ and $q_{\sscr{\pm}}$ denote
two independent solutions present at each instant of time $t\neq 0$. For $\o
<0$ one of them describes an expanding and the other one a collapsing
universe. For $t<0$ the expanding solution is the superinflationary branch.
All stochastic quantities depend on the mode functions of $\F$ which, for the
background given in~\refs{V0solutions}, read
\be
\vf_{\bk} (t) \propto |\, t\, |^{-\a(\o)}\, H^{(2)}_{\n(\o)}  
\left( \b(\o, |\bk |)\, |\, t\, |^{\g(\o)} \right) \ .
\label{modes1}
\ee
The wave vector is denoted by ${\bk}$ and $\a$, $\b$, $\g$ and $\n$ are
functions of $r_{\sscr{\pm}} (\o)$ or $q_{\sscr{\pm}} (\o)$. The choice of the
second Hankel function guarantees the correct Minkowski limit. Background
solutions can also be given for a linear potential $V(\F)=-2\l\F$, whereas the
corresponding mode functions are known only as a power
series~\cite{lalak-poppe}.

We split the field and its first time derivative in super-- and
subhorizon--sized parts,
\be
\F({\bf x}, t) = \F_{<} ({\bf x}, t) + \F_{>} ({\bf x}, t) 
\quad\mbox{and}\quad
\dot{\F}({\bf x}, t) = v_{<} ({\bf x}, t) + v_{>} ({\bf x}, t) \ ,
\label{split2}
\ee
containing all mode functions with physical wave vectors smaller ("$\scr{<}$")
and larger ("$\scr{>}$") than $aH$, respectively. It is not difficult to
derive a set of stochastic differential equations (Langevin equations)
describing the time evolution of the long wavelength parts (interpreted as the
semiclassical background henceforth) under the influence of two independent
stochastic noise terms. Solutions to these equations automatically incorporate
the backreactions. The fluctuation strength of the noise terms have to be
evaluated with the help of the two point correlation functions of the relevant
field operators. They are given in terms of the mode functions and the
background variables.

A numerical integration of the stochastic equations of motion for the
superinflationary pre--big--bang branch reveals the significant effect of the
backreactions: The dispersion of the induced fluctuations in the Hubble
parameter $H = \dot{a} / a$ grows much quicker for $t\ra 0_{-}$ than the
difference $\D_{H}$ separating the two distinct values of $H$ existing for all
times $t\neq 0$. The stochastic ensembles belonging to the expanding and the
collapsing solution of the classical theory overlap and quantum effects
dominate the dynamics. Therefore, there is no reason to continue the evolution
{\em after} the big bang with a preferred choice of one of the (semi)classical
solutions, unless one is able to make statements about the probabilities
(e.~g. by solving the corresponding Fokker--Planck equation) that one or the
other solution is preferentially occupied by the stochastic ensemble. Thus,
the system can emerge from the quantum regime as an ordinary universe with a
decelerated expansion rate, corresponding to one of the two post--big--bang
branches.

The qualitative behaviour remains unchanged if a linear potential is taken
into account, as long as $\l\lesssim\co (1)$. An analytical
treatment~\cite{lalak-poppe} is consistent with the numerical integration and
shows in addition the strong dependence of this effect on the JBD parameter
$\o$. For $\o\ra 0_{-}$ the backreactions get drastically suppressed, whereas
the opposite is true the more negative $\o$ gets. This fact originates from
the strong dependence of the Hubble parameter on $\o$.

\section*{Acknowledgments}
We thank the organisers, especially L.~Roszkowski, for their hospitality
during the workshop. This work was supported by Sonderforschungsbereich 375-95
"Research in Astroparticlephysics" of DFG, the European Commission TMR
programmes ERBFMRX-CT96-0045 and ERBFMRX-CT96-0090, and by the Polish Commitee
for Scientific Research grant 2 P03B 040 12.

\end{document}